\documentclass[letter,aps,pra,amsmath,amssymb,reprint,showkeys,showpacs]{revtex4-1} 

\usepackage{graphicx}
\usepackage{subfigure}

\newcommand{\unit}[1]{\ensuremath{\, \mathrm{#1}}}

\begin{document}

\title{Coherent manipulation of cold Rydberg atoms near the surface of an atom chip}
\author{J.~D.~Carter}
\author{J.~D.~D.~Martin}
\affiliation{Department of Physics and Astronomy and Institute for Quantum Computing, University of Waterloo, Waterloo, Ontario, Canada N2L 3G1}
\date{\today}

\begin{abstract}

Coherent superpositions of the $49s_{1/2}$ and $48s_{1/2}$ Rydberg states
of cold $^{87}$Rb atoms were studied near the surface of an atom
chip.  The superpositions were created and manipulated using
microwaves resonant with the two-photon $49s_{1/2}-48s_{1/2}$ transition.  Coherent behavior was observed using Rabi flopping,
Ramsey sequences, spin-echo and spin-locking.  These results are
discussed in the context of Rydberg atoms as electric field noise
sensors.  We consider the coherence of systems quadratically coupled
to noise fields with $1/f^{\kappa}$ power spectral densities ($\kappa \approx 1$).

\end{abstract}


\maketitle



\section{Introduction}

The coherent manipulation of atoms near surfaces is important for the potential development of hybrid quantum systems that will combine the advantages of solid-state devices with those of gas-phase atoms \cite{PhysRevA.79.040304,xiang:2013,saffman:2010}.  Treutlein {\it et al.}~\cite{treutlein:2004} have created
coherent superpositions of different hyperfine levels of $^{87}$Rb atoms trapped in an atom chip and measured $\approx 1 \unit{s}$ coherence times (limited by background magnetic field fluctuations).  In the present work, we study coherent superpositions of Rydberg states near an atom chip; these superpositions are highly susceptible to {\em electric} field fluctuations.

The development of both magnetic microtrapping (atom chips) and optical trapping of laser cooled atoms has recently allowed precise studies of Rydberg atom energy level shifts as a function of atom-surface distance \cite{tauschinsky:2010,hattermann:2012} (the shifts are primarily due to adsorbate fields).  These studies used electromagnetically-induced transparency as a probe of Rydberg level shifts.   In contrast, the {\em selective} field ionization technique (SFI) \cite{gallagher:1994} allows the populations of different Rydberg levels to be measured, and is ideal for the study of Rydberg atom coherences.   For example, Hogan {\it et al.}~\cite{hogan:2012} were able drive coherent Rabi oscillations between two Rydberg states of helium near a co-planar waveguide, selectively detecting the states using SFI.

Our laboratory recently demonstrated the SFI of Rydberg atoms in proximity to an atom chip \cite{carter:2012}, and characterized energy level shifts associated with electric fields due to dielectric charging.  In the present work, we create, manipulate, and measure Rydberg coherences near the chip using resonant microwave fields and SFI.  We find that the Rydberg coherence times can be studied and extended using techniques from the ubiquitous ``two-state toolbox'' \cite{vandersypen:2005}: Ramsey interferometry, spin-echo, and spin-locking.  Additionally, we find that the decay of Rydberg atom coherences is a useful probe of electric field fluctations near surfaces.

.

\section{Experimental techniques}

To determine the influence of the atom chip surface on Rydberg state coherence we have compared two scenarios : 1) Rydberg excitation of cold atoms after release from an atom chip, at a distance of $150 \unit{\mu m}$ from the chip surface, and 2) Rydberg excitation after release of cold atoms from a magneto-optical trap (MOT) approximately $3\unit{mm}$ away from the surface.

\begin{figure}
\includegraphics{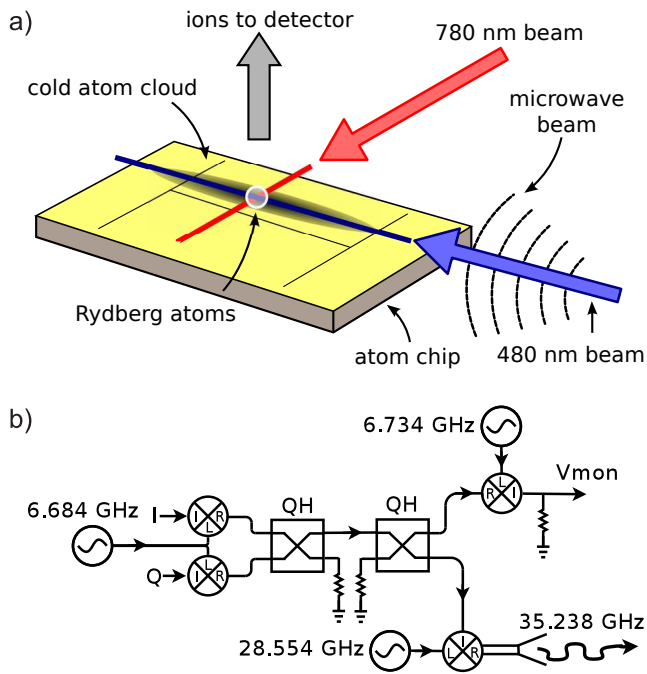}
\caption{\label{fg:apparatus} (Color online) (a) Schematic illustrating the Rydberg excitation geometry.  This illustration is upside down --- the atoms are below the chip in the experimental apparatus.  (b) Schematic of microwave modulation system.  QH: quadrature hybrid.  The monitor output ${\rm V_{mon}}$ is useful for correcting modulation imperfections \cite{nash:2009}.  It is used to determine the IQ dc offset corrections required to minimize the IQ modulator LO leakage, and in the case of multiple pulses, for adjusting the I and Q inputs to obtain accurate relative phases for the different pulses.}  
\end{figure}

The protocol for release from the atom chip is similar to that presented in Ref.~\cite{carter:2012}.  In particular, atoms are collected in vapor cell MOT, cooled in optical molasses, optically pumped into the $5s_{1/2},F=2,m_F=2$ sublevel, and then magnetically trapped in a macroscopic Ioffe-Pritchard trap.  From this trap the atoms are transferred to a wire-based magnetic microtrap \cite{fortagh:2007,reichel2010atom}, positioned, and then the wire currents are shut off --- the magnetic field orientation and magnitude change slowly enough so that $m_F$ is preserved.  The atoms are left in a homogeneous $34.5\unit{G}$ magnetic field.   Excitation to Rydberg states is a two-step process: a $780\unit{nm}$ beam is used to excite from the $5s_{1/2},F=2,m_F=2$ state to $5p_{1/2},F=3,m_F=3$, and a $480\unit{nm}$ beam excites the $5p_{1/2},F=3,m_F=3 \rightarrow 49s_{1/2}, m_j=1/2$ transition.   The $480\unit{nm}$ light is obtained from a frequency doubled Ti:sapphire laser, frequency stabilized using a transfer cavity \cite{cels2:2011,*bohlouli:2006}.  The $480\unit{nm}$ beam runs in the same direction as the wires (the ``long'' dimension of trap) and defines the distance to the surface: $150 \pm 30 \unit{\mu m}$.  A $780\unit{nm}$ beam runs perpendicular to the $480\unit{nm}$ beam, restricting the length of the Rydberg sample to approximately  $1.2\unit{mm}$ (see Fig.~\ref{fg:apparatus}).
The $480\unit{nm}$ beam is vertical polarized.  The $780\unit{nm}$ excitation beam is circularly polarized and propagates in the direction of the $34.5\unit{G}$ magnetic field.
 
In previous work on the optical Rydberg excitation spectra near the atom chip surface \cite{carter:2012} we found that the electric fields near the chip depended on how the chip wire currents had been applied, even though the wire currents were switched off prior to excitation.   In particular, if while current was flowing the wires had a voltage bias (due to ohmic drops) that was positive with respect to nearby grounded surfaces, the electric fields were much larger than if the bias was negative.  In the present work we exclusively use the lower-field negative bias situation.  By detuning the $480\unit{nm}$ source to the blue side of the optical excitation peak, and scanning the voltage on the metal plates below the chip surface for maximum Rydberg signal, we can determine the ``compensating'' plate voltage that produces a net average zero electric field over the Rydberg excitation volume (see Fig.~2(b) of Ref.~\cite{carter:2012}).  All experiments in this work are done in this compensated configuration.

Coherent superpositions of $49s_{1/2}$ and $48s_{1/2}$ were created and manipulated using pulsed microwaves resonant with the two-photon $49s_{1/2}-48s_{1/2}$ transition---the resonant microwave frequency is $\approx 35.238\unit{GHz}$ \cite{li:2003}.   The amplitude, duration and relative phase of the pulses were precisely controlled using a IQ modulator based system [see Fig.~\ref{fg:apparatus}(b)].  A $6.684\unit{GHz}$ fixed frequency source drives the local oscillator (LO) input of an in-phase/quadrature (IQ) modulator. The pulse modulation at its output (RF) is controlled through the I and Q inputs, sourced using a dual DAC (200 MSamples/s/channel) driven by a field-programmable-gate-array (FPGA).  Pulse sequences are downloaded from a computer to the FPGA using a USB-UART (universal asynchronous receiver/transmitter) interface.  The modulated $6.684\unit{GHz}$ is applied to the intermediate frequency (IF) port of a mm-wave mixer (Wisewave FUB-28-01), with the LO sourced by a tunable synthesizer ($\approx 28.554\unit{GHz}$).  The upper sideband is resonant with the $49s_{1/2}-48s_{1/2}$ transition ($\approx 35.238\unit{GHz}$).  The rf output port of the mixer is a WR28 waveguide connected to a horn aimed towards the atoms in the vacuum chamber. 

Following microwave manipulation a ramped electric field is applied using macroscopic conducting plates located $\approx 2\unit{cm}$ away from the chip surface (see Fig.~1(a) of Ref.~\cite{carter:2012}; these are the same plates used to apply the compensating field voltage mentioned above).  The $48s_{1/2}$ and $49s_{1/2}$ states are ionized at different fields, and the resulting ions appear at different times in the microchannel plate charged particle signal.  This selective field ionization signal (SFI) allows us to measure the individual state populations.  We integrate signals over the times associated with each of these signals, and normalize the $48s_{1/2}$ signal by the total $48s_{1/2}$ and $49s_{1/2}$ signal.  In all of our results there is a slight uncorrected bias towards preferential detection of $49s_{1/2}$ states over $48s_{1/2}$, leading to what should be a 50\% mixture of $49s_{1/2}$ and $48s_{1/2}$ being detected as a $48s_{1/2}$ fraction of $\approx 45$\% (varying by a few percent from day to day).  At least some of this bias is caused by a small fraction of $48s_{1/2}$ states being ionized at approximately the same electric field as $49s_{1/2}$ (i.e.~at earlier times in the SFI ramp).


\section{Results}

Rabi flopping is the canonical coherent manipulation of a two-state system:  an oscillating field resonantly couples two energy eigenstates.  With the population initially in one of the two states, it ``flops'' back and forth between the two states at a rate dependent on the coupling.  For an ensemble this oscillation can decay for a variety of reasons, including:  1) spatial inhomogeneity in the driving field, giving a distribution of Rabi frequencies, 2) spatial inhomogeneity in the energy levels over the sample, and 3) time-dependent perturbations of the energy levels, leading to decoherence.  Rabi flopping has been observed with many disparate physical systems: in nuclear and electron magnetic resonance, superconducting qubits \cite{Vion_Sci.03052002}, and microwave transitions of Rydberg atoms \cite{gentile:1989}, to name but a few.

\begin{figure}
\includegraphics{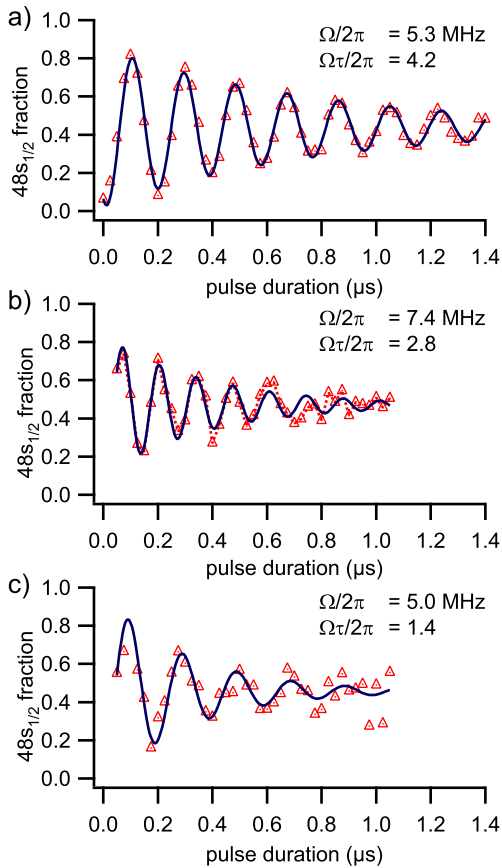}
\caption{\label{fg:rabi_flopping} (Color online)
Rabi flopping of initially excited $49s_{1/2}$ states as a function of microwave pulse duration (a) after release from MOT, and (b) after release from microtrap with the $780\unit{nm}$ excitation beam centered on standing wave maximum of microwaves, and (c) displaced 1 mm from the maximum.  
The difference in the Rabi frequencies of (b) and (c) illustrates the spatial inhomogeneity of the rf.}
\end{figure}

Figure \ref{fg:rabi_flopping} illustrates Rabi flopping in our system.  The initially excited $49s_{1/2}$ state is resonantly coupled to $48s_{1/2}$.  The $48s_{1/2}$ fraction oscillates, eventually decaying to approximately 50\% of the total Rydberg population.  Despite the restricted volume the Rabi oscillations damp out more quickly near the chip surface [compare Fig.~\ref{fg:rabi_flopping}(a) with (b) and (c)].  The significant change in Rabi frequency as the Rydberg sample position is changed (using the $780\unit{nm}$ excitation laser alignment) suggests that spatial inhomogeneity in the driving field causes the Rabi oscillations to damp near the chip.  This is supported by the observation that the time constant for the decay of the Rabi oscillations $\tau$ scales reliably with the Rabi frequency $\Omega/2\pi$, such that $\tau \Omega/2\pi$ remains constant over a range of $\Omega/2\pi$.  Also the Rabi oscillations damp out more quickly with a larger $780\unit{nm}$ beam diameter.   The increase in microwave inhomogeneity near the chip is most likely due to the discontinuities presented by the chip surface i.e.~the chip edges will cause microwave reflections, helping support partial standing waves.

A Ramsey pulse sequence can allow us to reduce the role of driving field inhomogeneity:  an initial $\pi/2$ pulse can be used to create a coherent superposition of the $49s_{1/2}$ and $48s_{1/2}$ states.  Following a free evolution waiting period, the superposition can then be probed by a second $\pi/2$ pulse.  The waiting time is varied to study the dephasing and/or decoherence during free evolution, while reducing the influence of pulse inhomogeneities.  To visualize the expected results for this and subsequent manipulations it is useful to invoke the Bloch sphere picture (see, for example, Ref.~\cite{vandersypen:2005}):  the first $\pi/2$ pulse rotates the Bloch vector from $+z$ into the $xy$ plane.  In the absence of decay the Bloch vector will then rotate in the $xy$ plane by a rate dependent on the detuning between the applied rf and actual two-level Larmor frequency.  With no detuning and a final $\pi/2$ pulse with the same initial rf phase as the first, the population should all be moved out of the initially populated state ($49s_{1/2}$ in our case).

The influence of resonant pulses can be considered as rotations of the Bloch vector, with the axes of rotation in the $xy$ plane.  The rf phase controls where the axis of rotation lies in the $xy$ plane.  Since this is a two-photon transition an rf phase shift of $\phi$ will change the rotation axis by $2\phi$.  To be more specific, in this work the phase of the initial rf pulse defines the $x$-axis (it is a $\pi/2$ rotation about the $x$-axis); further pulses with their rf shifted by $45^\circ$ will rotate around the $y$-axis of the Bloch sphere. Analogous behavior for the phase shift in 2-photon transitions has been observed in NMR \cite{hatanaka:1978}.

\begin{figure}
\includegraphics{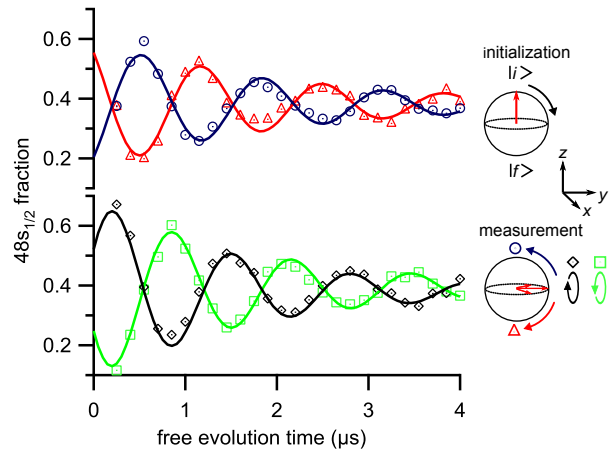}
\caption{\label{fg:ramseyphases} (Color online)
Ramsey sequences for atoms released from the MOT, initially excited to $49s_{1/2}$, as a function of free evolution time between two $\pi/2$ pulses ($50\unit{ns}$ each), with the two-photon energy of the rf detuned by $+1\unit{MHz}$; the rf detuning is $+0.5\unit{MHz}$.  The different cases illustrate the role of the rf phase difference between the two $\pi/2$ pulses of a Ramsey sequence.}
\end{figure}

Figure \ref{fg:ramseyphases} illustrates the Ramsey sequence and our control of the rf phase using detuned rf.  Once rotated into the $xy$ plane, the average Bloch vector precesses due to the detuning.  By applying $\pi/2$ measurement pulses about both the $x$ and $y$-axis, we can observe this precession.  When a rotation about $x$ gives a maximum final $48s_{1/2}$ signal, a $y$ rotation gives a 50\% mixture of $48s_{1/2}$ and $49s_{1/2}$, and vice versa. 

The Ramsey sequence of Fig.~\ref{fg:ramseyphases} (and all other sequences reported here) use $50\unit{ns}$ duration $\pi/2$ pulses.  To set the rf levels for $\pi/2$ pulses we optimize for maximum transfer from the initially excited $49s_{1/2}$ to $48s_{1/2}$ for a sequence consisting of two $50\unit{ns}$ pulses separated by $50\unit{ns}$.  To within our measurement uncertainty, the $100\unit{ns}$ duration $\pi$ pulses we use later also require the same amplitude.

\begin{figure*} 
\includegraphics{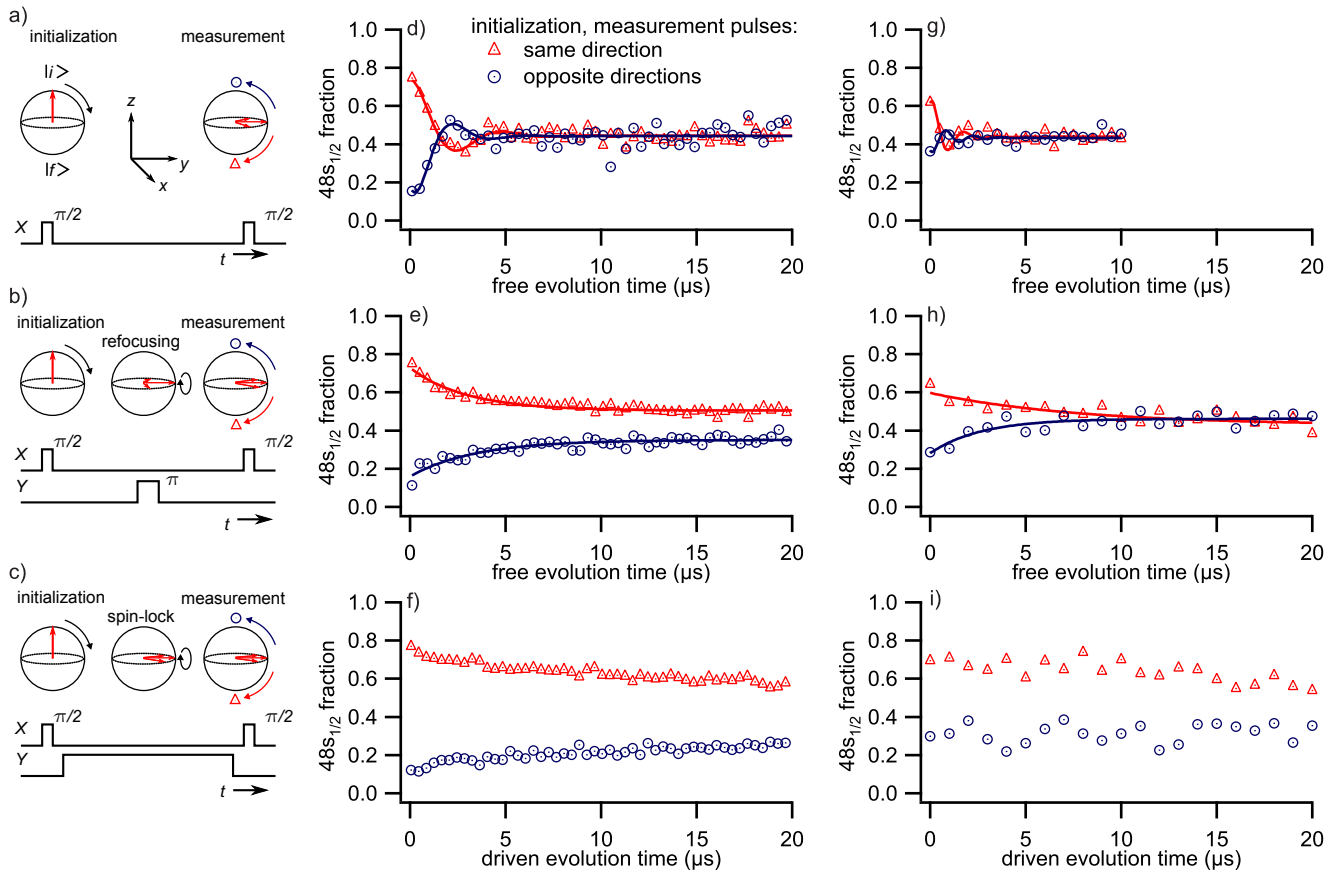}
\caption{\label{fg:big_plots} (Color online)
(a-c) Control sequences for (a) Ramsey, (b) spin-echo, and (c) spin-locking. The initialization pulse rotates the Bloch vector by an angle of $\pi/2$ around the $x-$axis to create a coherent superposition of the $49s_{1/2}$ and $48s_{1/2}$ Rydberg states. The measurement pulse rotates the Bloch vector by an angle of $\pi/2$ around an axis lying in the equator; for the measurements shown in this figure the initialization and measurement pulses are along the same axis. In the spin-echo sequence, coherence is with a refocusing $\pi-$pulse as shown in (b), the spin-locking sequence requires continuous rotation around the $y-$axis as shown in (c). (d-f) Time-evolution of coherence for atoms released from the MOT, $\approx 3 \unit{mm}$ from the surface. Control sequences are (d) Ramsey, (e) spin-echo, and (f) spin-locking. (g-i): time-evolution of coherence for atoms released from the microtrap, $\approx 150 \unit{\mu m}$ from the surface. Control sequences are (g) Ramsey, (h) spin-echo, and (i) spin-locking. In (d-i), red traces are experiments with identical initialization and measurement pulses, and blue traces are experiments where the initialization and measurement pulses rotate along the same axis in opposite directions.}
\end{figure*}

In Fig.~\ref{fg:big_plots}(d) and (g) Ramsey sequences from both MOT release and chip release are shown.  The rapid damping of the chip case is expected due to larger static electric field inhomogeneities \cite{carter:2012}.
The rf frequency for the $\pi/2$ pulses was determined by weak field spectroscopy of the $49s_{1/2}-48s_{1/2}$ line.  The Ramsey results of Fig.~\ref{fg:big_plots} exhibit oscillation with a period on the order of the damping time.  This oscillation can be removed by shifting the rf to a slightly higher frequency.

The coherence measured by a Ramsey sequence will be susceptible to inhomogeneities in the energies of the atoms over the sample.  The classic approach to dephasing due to spatial inhomogeneities --- discovered in the early days of NMR --- is spin-echo \cite{hahn:1950}:  after an initial $\pi/2$ pulse the Bloch vectors (each representing different Larmor frequencies in the sample) will spread apart.  After an initial time period a ``refocusing'' $\pi$ pulse can be applied, which rotates by $\pi$ around either the $x$ or $y$-axis, depending on the rf phase.  If the precession of each Bloch vector around the $y$-axis is constant in time, the sample will ``rephase'' and all the Bloch vectors meet again a time later, equal to the time between the initial $\pi/2$ preparation pulse, and $\pi$ refocusing pulse.    As Fig.~\ref{fg:big_plots}(b) shows, this procedure increases the free-evolution time over which coherence is preserved compared to a Ramsey sequence.   Analogous refocusing behavior has been observed using relatively slow dc field modulation for coherent superpositions of Rydberg states \cite{minns:2006,yoshida:2007,yoshida:2008}, and in Rydberg optical excitation \cite{raitzsch:2008}.

\begin{figure}
\includegraphics{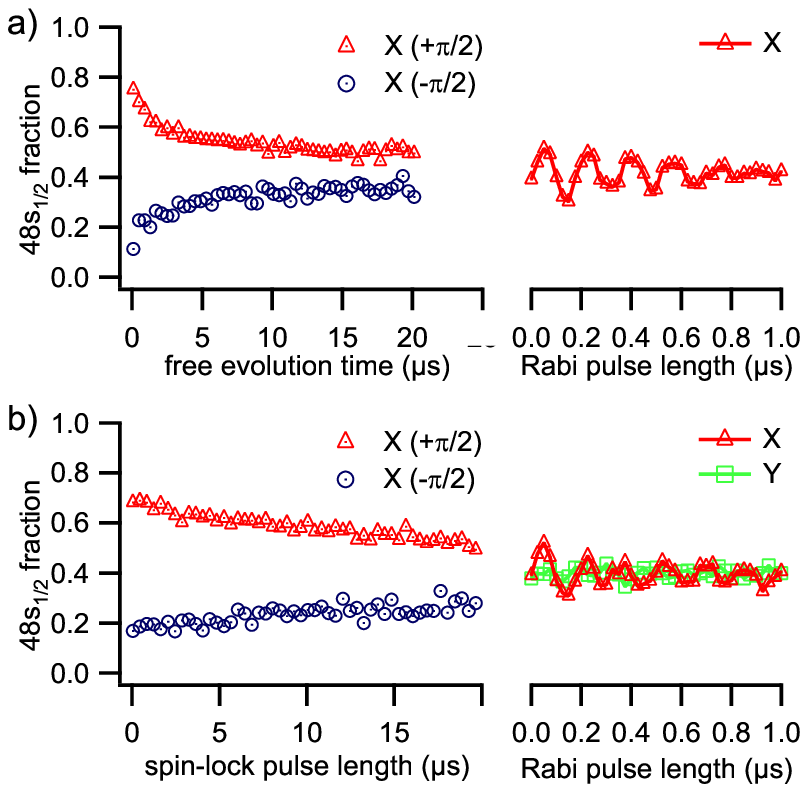}
\caption{\label{fg:rabi_after} (Color online)
Rabi flopping (right hand panels) as a function of the duration of a resonant pulse, observed at the end of (a) a spin-echo sequence of $20\unit{\mu s}$ free-evolution time and (b) a $20\unit{\mu s}$ duration spin-locking pulse sequence (both after release from a MOT).  The left hand panels indicate the populations observed at the end of the sequences with no Rabi flopping pulse applied.
As expected, in (b) the correct rf phase leads to Rabi flopping of comparable initial amplitude to the measured coherence at the end of the spin-echo sequence, whereas rf shifted by $45^{\circ}$ (which rotates about the $y$-axis) leads to no detectable oscillations. }
\end{figure}

The differences between the detected $48s_{1/2}$ fractions under alternating $\pm \pi/2$ pulses about orthogonal axes in the $xy$ plane is a good measurement of coherence in this system, where relaxation between $48s_{1/2}$ and $49s_{1/2}$ does not have to be considered.  However, as an additional verification of the preservation of coherence we can also initiate Rabi flopping at the end of a spin-echo sequence (see Fig.~\ref{fg:rabi_after}(a).  

If spin-echo and spin-locking ({\it vide infra}) are performed on atoms released from the MOT at lower Rydberg density than Fig.~\ref{fg:big_plots} the observed coherence improves; the fast initial decay of the spin-echo (Fig.~\ref{fg:big_plots}(e) disappears, and the long-time decay constants are significantly longer than $20\mu s$.  This points to Rydberg atom interactions as a source of decoherence.  However as the MOT and chip release results of Fig.~\ref{fg:big_plots} were taken at comparable Rydberg density, it is reasonable to conclude that the excess decay of Fig.~\ref{fg:big_plots}(h)  over (e) is due to proximity to the chip, and an increase in electric field noise.

The eventual decay of coherence near the chip, even with a refocusing
pulse, suggests the use of multiple refocusing pulses to preserve coherence \cite{viola:1998}.  Unfortunately, our large rf inhomogeneity is expected
to limit the effectiveness of this approach (at least the simplest
implementation).  Instead we have pursued the idea of
``spin-locking'' \cite{redfield:1955} (as with spin-echo, our usage of the term ``spin'' is figurative, and refers to the analogy between the two Rydberg states and a spin-1/2 system in a magnetic field).  The experimental procedure for spin-locking is similar to a Ramsey experiment, in that two $\pi/2$ rotations are used to create and measure a coherent superposition, separated by a time interval in which the coherent superposition evolves. However, in the case of a spin-locking experiment, the evolution during this interval is driven evolution, with a continuous rotation around the axis of the Bloch vector, rather than the free evolution used in the Ramsey case (the driven evolution duration is typically much larger than the initial or final $\pi/2$ pulse durations; see Fig.~\ref{fg:big_plots}(c). Specifically for our convention where the initial $\pi/2$ pulse rotates about the $x$-axis leaving the Bloch vector pointing in the $y$-direction, the spin-locking pulse rotates about the $y$-axis, having no effect in a system without decoherence or dephasing.  After the spin-locking pulse, a final $\pi/2$ measurement pulse rotates about $x$ before the final state populations are measured.

Decoherence during a spin-locking sequence is dominated by
noise near the Rabi frequency of the driven evolution.  The influence of low frequency noise and inhomogeneities is suppressed.  The effectiveness of driven evolution for reducing Rydberg state decoherence has been demonstrated by Minns {\it et al.} \cite{minns:2008}.

As Fig.~\ref{fg:big_plots}(c) illustrates, the spin-locking pulses preserve coherence, and offer an improvement over spin-echo close to the chip.  The dramatic contrast between spin-locking and the full decay of coherence observed for spin-echo points to the frequency dependence of the noise ({\it vide infra}).   As with spin-echo, we can confirm the coherence left at the end of the sequence by performing a Rabi oscillation experiment [see Fig.~\ref{fg:rabi_after}(b)].

One appeal of spin-locking is the systematic way in which the loss of coherence depends on the spectral noise density.  In a system with linear coupling to the noise field, the decoherence rate is simply related to the noise spectral density at the Rabi frequency corresponding to the spin-lock pulse (see, for example Ref.~\cite{loretz:2013}).   However as we have nulled out the dc electric field, the electric field $F$ coupling is quadratic, due to the polarizability difference between the $49s_{1/2}$ and $48s_{1/2}$ states \cite{jones:2013}.  In this case, spin-locking is sensitive to the statistics of $F^2$, not $F$, and thus noise in $F$ at pairs of frequencies $f_1$ and $f_2$ can contribute to the noise in $F^2$ at $\Omega/2\pi$, when $\Omega/2\pi=f_1+f_2$, $f_1-f_2$, etc...  This is a significant complication, and for noise spectroscopy of Rydberg atoms it is almost certainly preferable to induce linear coupling with a dc bias field (see Section \ref{se:limits}).  Nonetheless, since noise spectral densities typically scale like $1/f^{\kappa}$ (with $\kappa$ ranging from $1/2$ to $2$), we expect spin-locking sequences to exhibit better preservation of coherence with higher Rabi frequencies for the spin-lock pulse.

\begin{figure}
\includegraphics{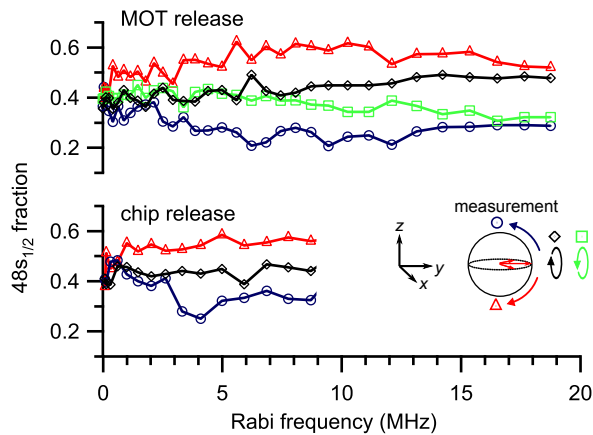}
\caption{\label{fg:spin_lock_amplitude} (Color online)
Populations measured after a $\pi/2$ initialization pulse, $20\unit{\mu s}$ long spin-lock pulse, and various $\pi/2$ measurement pulses (inset), as a function of the Rabi frequency of the spin-lock pulse.
}
\end{figure}

Figure \ref{fg:spin_lock_amplitude} indicates the coherence left after spin-locking pulses of varying strength, with this strength measured in terms of the Rabi frequency of the spin-lock pulse (the spin-locking results of Figs.~\ref{fg:big_plots}(f), (i) and ~\ref{fg:rabi_after}(b) use 5 MHz Rabi frequency spin-lock pulses).  Starting from a weak spin-locking pulse amplitude and increasing the Rabi-frequency improves the coherence left at the end of the sequence, as one would expect for a system with predominantly low-frequency noise.  However, in the MOT case (studied to higher Rabi frequencies than in the microtrap release case) the measurement along the $y$-axis eventually shows reduced contrast at higher frequencies, with an increase in contrast appearing along the $x$-axis.  This is possibly due to a power-dependent phase shift in the final stage mm-wave mixer (see Fig.~\ref{fg:apparatus}(b). Such a phase shift would result in the spin-lock axis of rotation not being aligned along the $y$-axis, but instead slightly rotated in the $xy$-plane.   (These phase shifts do not influence the results of Fig.~\ref{fg:ramseyphases}, as the two Ramsey $\pi/2$ pulses have equal amplitude.)

\section{Noise and Rydberg atom coherence}
\subsection{Introduction}
The experimental results of the previous section will now be discussed in the context of using Rydberg atoms as electric field noise sensors.  The decoherence that we have observed may be useful in the characterization (and elimination) of field noise during the development of both hybrid quantum devices \cite{PhysRevA.79.040304,xiang:2013,saffman:2010} and Rydberg gates near surfaces \cite{muller:2011}.  An additional motivation for undertaking the measurement of electric field noise near surfaces is the ``anomalous'' heating in ion traps~\cite{Turchette_PRA.61.063418,Wineland_PRL.109.103001,Haffner_NJP.13.013032}, caused by field noise with a roughly $1/f$ frequency dependence near electrode surfaces.  The frequency range of interest for ion traps is typically in the MHz regime, making coherent superpositions of Rydberg atoms more useful for the study of this noise when compared to electric field sensing using {\em resonant} Rydberg transitions \cite{bradley:2003,sedlacek:2013} at higher frequencies.

It has been recognized that spin-echo sequences act as a bandpass frequency filter on environmental noise~\cite{bylander:2011}, so that by systematically changing the decoupling sequence and observing the effect on decoherence it is possible to measure the power spectral density of the noise, which can help to determine the noise source. In our Rydberg atom experiments, loss of coherence can be inferred from the relative populations of the two states following the measurement pulse (assuming the rotation axis is perpendicular to the Bloch vector):
\begin{equation}
\label{eq:population_dephase}
| \langle e^{i \delta \phi(t)} \rangle | 
=\frac{p(f)-p(i)}{p(f)+p(i)}.
\end{equation}

Two processes contribute to the loss of coherence. Relaxation of the populations of $|i\rangle$ and $|f\rangle$ toward thermal equilibrium cause longitudinal relaxation, which is exponential in time with a rate $\Gamma_1=1/T_1$. In Rydberg atoms the dominant process is usually population transfer caused by coupling to blackbody radiation from the surrounding environment. Losses to states outside the two-level system do not contribute to $\Gamma_1$ \cite{minns:2006}, but ultimately limit the practical measurement time.

The second process is dephasing, caused by fluctuations in the energy separation of $|i\rangle$ and $|f\rangle$.  In our case this is due to the polarizability difference between the $49s_{1/2}$ and $48s_{1/2}$ states of Rb \cite{jones:2013}.  Dephasing is often characterized by a time constant $T_2$ but is not necessarily exponential in time, particularly if the noise power spectrum diverges at $\omega=0$ \cite{Ithier_PRB.72.134519}. We write the contribution of pure dephasing as $| \langle \exp[i \delta \phi(t)]\rangle | = f_z(t)$, and the total rate of coherence decay takes the form $\exp[-\Gamma_1 t]f_z(t)$~\cite{Ithier_PRB.72.134519}. In the $49s_{1/2}-48s_{1/2}$ system discussed in this paper, relaxation between the two states is negligible compared to the rate of dephasing, so $\exp[-\Gamma_1 t] \approx 1$.

\subsection{Free evolution}
With short refocusing pulses and the energy level difference $\hbar \omega_{if}$ depending linearly on $\lambda$ (a generic external field), $f_z(t)$ can be computed using \cite{Ithier_PRB.72.134519}:
\begin{equation}
\label{eq:coherence_linear}
f_z(t) = \left \langle \exp\left[i \frac{\partial \omega_{if}}{\partial \lambda}\int_0^t d\tau  G(\tau) \lambda(\tau) \right] \right \rangle,
\end{equation}
where $G(\tau)$ is a gating function determined by the timing of the refocusing pulses. During free evolution, $G(\tau)=\pm 1$, with the sign changing at each $\pi$ pulse (pulses are assumed to be infinitesimally short). For example, a Ramsey experiment (with no refocusing pulses) will simply have $G(\tau)=1$ at all times, and a single pulse spin-echo experiment with total free evolution time $t$ will have $G(\tau<t/2)=+1$ and $G(\tau>t/2)=-1$.

If the statistics of $\lambda$ are Gaussian and its power spectral density is $S_\lambda(\omega)$, the dephasing is given by Eq.~(2) in Bylander \emph{et al.}~\cite{bylander:2011}:
\begin{equation}
\label{eq:coherence_integral}
f_z(t)=\exp\left[-\tau^2\left(\frac{\partial\omega_{if}}{\partial\lambda}\right)^2\int_0^\infty d\omega S_\lambda(\omega)g_N(\omega,\tau)\right].
\end{equation}
Here, $\tau$ is the total free evolution time of the superposition, and $N$ is the number of $\pi$-pulses. The filter function $g_N$ depends on the number and timing of the $\pi$-pulses, and is given by (cf. Eq.~(3) in Ref.~\cite{bylander:2011})
\begin{eqnarray}
\label{eq:filter_function}
g_N(\omega,\tau)=\frac{1}{(\omega \tau)^2} \left| 1 + (-1)^{1+N}e^{i\omega \tau} \right. \\
+  2 \sum_{j=1}^N \left.(-1)^j e^{i\omega \delta_j \tau} \cos(\omega \tau_\pi/2)\right|^2,
\end{eqnarray}
where $\tau_\pi$ is the length of each $\pi$-pulse. Therefore, the total length of the pulse sequence is $\tau +N \tau_\pi$. The normalized position of the center of the $j$th $\pi$-pulse is $\delta_j=t_j/(\tau +N \tau_\pi)$.

If the refocusing pulses are evenly spaced, $g_N(\omega,\tau)$ in Eq.~\ref{eq:filter_function} will peak at $\omega_N=\pi N /T$, and the coherence integral of  Eq.~\ref{eq:coherence_integral} is approximately
\begin{equation}
\label{eq:approx_coherence_integral}
f_z(t)\approx \exp\left[- \tau \left(\frac{\partial \omega_{if}}{\partial \lambda} \right)^2 S_\lambda(\omega_N)\right],
\end{equation}
assuming $S_\lambda(\omega)$ is reasonably constant over the filter function's bandwidth of $\approx 2 \pi /\tau$.

If the shift in $\omega_{if}$ is linear with electric field $F$, then writing the dephasing in terms of $S_F(\omega)$ is straightforward---$F$ is simply the noise source $\lambda$. This linearity may be achieved for optically accessible Rydberg states --- of predominantly low angular momentum --- by application of a static dc field (large compared to the rms field noise).  This ``biasing'' has the additional advantage of increasing the sensitivity of $\omega_{if}$ to changes in $F$.

Noise measurement pulse sequences have been used, for example, to measure diffusion through porous media in NMR~\cite{Lasic.JMagRes.182.208} and noise in superconducting qubits~\cite{Ithier_PRB.72.134519,bylander:2011}.
Pulsed refocusing sequences have extended coherence times of Rydberg atoms~\cite{minns:2006,yoshida:2008}, but have not, so far, been used to measure environmental noise.

If the leading order of the Stark shift is quadratic (i.e.~the bias field is zero), higher orders of the statistics of $F$ become important, and analytical solutions for the evolution of $f_z(t)$ are not always feasible. We discuss numerical calculations of $f_z(t)$ for $1/f$-type noise in the Appendix.
 
\subsection{Driven evolution}
Measuring high-frequency noise with pulsed-decoupling sequences may require undesirably high modulation bandwidths or microwave power in order to generate sufficiently short pulses. In such situations, spin-locking \cite{redfield:1955,Ithier_PRB.72.134519,loretz:2013} is a more desirable approach.

During driven evolution at Rabi frequency $\Omega$, the rotating-frame eigenstates $|\tilde{i}\rangle$ and $|\tilde{f}\rangle$ have energy splitting $\hbar \Omega$, and relax exponentially towards equal populations of $|\tilde{i}\rangle$ and $|\tilde{f}\rangle$ (assuming $\hbar \Omega \ll k_BT$). This relaxation is partly due to $T_1$ relaxation of the energy eigenstate $|i\rangle$ and $|f\rangle$ populations in the non-rotating frame, with an additional component due to energy fluctuations at the frequency $\Omega/(2\pi)$. At long times $t > 2/\tilde{\Gamma}_1$, the total relaxation rate (Eq.~42 in Ref.~\cite{Ithier_PRB.72.134519}) is
\begin{equation}
\tilde{\Gamma}_1= \Gamma_\nu \sin^2\eta+\frac{1 + \cos^2\eta}{2}\Gamma_1,
\end{equation}
where $\Gamma_\nu=\pi S_{\delta\omega_{if}}(\Omega)$ is related to the power spectral density of the fluctuations of the energy level separation $\omega_{if}$ at the Rabi frequency and $\eta$ is the angle between the axis of rotation and the $z$-axis, determined by the detuning $\Delta$ of the microwaves from resonance by $\Delta = \Omega \cos \eta$. Sensitivity to field noise is maximized when the axis of rotation lies along the equator of the Bloch sphere, i.e., $\eta=\pi/2$, which requires the applied microwaves to be resonant. 

Spin locking of a solid state qubit has been used to measure the noise in the phase $\delta$ across the Josephson junctions of a quantronium circuit at its optimal working point, where the lowest order of the coupling is quadratic in $\delta$~\cite{Ithier_PRB.72.134519} (see also Ref.~\cite{yan:2013}). Spin locking of the electron spin of a single nitrogen-vacancy center in diamond has also been recently used to sensitively detect MHz-frequency oscillating magnetic fields, where the lowest order of the coupling was linear~\cite{loretz:2013}.

The decay of Rabi oscillations may also be used to measure phase noise; in addition to the relaxation described above, there is also pure dephasing of the coherence between $|\tilde{i}\rangle$ and $|\tilde{f}\rangle$. This decay has been used to measure the $1/f$ flux noise in persistent-current qubits~\cite{bylander:2011} as well as noise in the phase of the Josephson junctions of a quantronium circuit at its optimal working point~\cite{Ithier_PRB.72.134519}, where measurements of the noise using Rabi flopping were consistent with spin-locking results. In our experiment, however, inhomogeneity of the microwave power across the size of the sample creates an additional source of dephasing which dominates the other decoherence mechanisms, making noise measurements using Rabi flopping impractical.

\subsection{Limits to sensitivity}
\label{se:limits}
The minimum detectable $S_F(\omega)$ depends on the polarizability of the Rydberg states involved. Maximum sensitivity is achieved with application of a dc electric field which is large compared to the magnitude of the field noise---this also makes the dominant order of the Stark shift linear, which simplifies determination of the noise spectrum.

Reasonably detectable dephasing has $|\ln f_z(t)| \gtrapprox 1$, so the minimum detectable field noise for a pulsed sequence may be estimated from Eq.~\ref{eq:approx_coherence_integral}:
\begin{equation}
\label{eq:Smin_pulse_taulimit}
S_{Fmin}(\omega_N)=\frac{1}{\tau (\partial \omega_{if}/\partial F)^2}.
\end{equation}

Similar considerations for spin-locking with a driven evolution time $\tau$ require $\Gamma_\nu \geq 1/\tau$, so that the minimum detectable field noise at the Rabi frequency $\Omega$ is 
\begin{equation}
\label{eq:Smin_spinlock}
S_{Fmin}(\Omega)=\frac{1}{\pi \tau(\partial \omega_{if}/\partial F)^2},
\end{equation}
which differs by a factor of $\pi$ from the pulsed-refocusing case for the same measurement time.

Ultimately, the maximum value of $\tau$ is limited by the effective lifetime of the Rydberg states, $\tau_{\rm eff}$ (limited by both spontaneous emission and blackbody radiation-induced transfer to nearby Rydberg states). In the case of pulsed-refocusing sequences, the maximum achievable $\tau$ for a given $\omega_N$ may be further limited by microwave inhomogeneities which restrict the maximum practical number of refocusing pulses.

One may increase both $(\partial \omega_{if}/\partial F)$ and $\tau_{\rm eff}$ by increasing the principal quantum number $n$. The maximum $\partial \omega_{if}/\partial F$ at a given $n$ is limited by the permanent dipole moment of the extreme Stark manifold states, with \cite{gallagher:1994} 
\begin{equation}
\label{eq:dipole_moment}
\frac{\partial \omega_{if}}{\partial F} \approx \frac{e a_0 n^{*2}}{\hbar}.
\end{equation}
The $n$ dependence of $\tau_{\rm eff}$ is complicated, involving contributions from both spontaneous emission and blackbody transfer (see for example Ref.~\cite{gallagher:1994}).  Nonetheless, like the permanent dipole moments, the effective lifetime $\tau_{\rm eff}$ also increases with $n$.

Several factors limit the sensitivity available in practice. Stark shifts due to uncompensated inhomogeneous dc electric fields (calculated from Eq.~\ref{eq:dipole_moment}) must be small compared to the Rabi frequency of any microwave manipulation. Energy level spacings scale as $1/n^3$, so at large $n$ maintaining large Rabi frequencies may not be possible without excitation of states nearby in energy. Finally, the matrix elements for optical excitation of Rydberg states drop sharply with increasing $n$, so at very large $n$ the Rydberg population may be too small for good signal/noise in the detection.

As an example of practically achievable sensitivity, we consider noise measurements in the $1-10 \unit{MHz}$ range, using the $50 s_{1/2}$ and $50 p_{3/2}$ states of Rb. Measurement lifetime is limited by $\tau_{\rm eff}=65 \unit{\mu s}$ for the $50 s_{1/2}$ state \cite{beterov:2009}. Maximum practical Rabi frequencies for pulses are $\Omega\approx 2 \pi \times 100 \unit{MHz}$, limited by undesired excitation to the nearby $50 p_{1/2}$ state ($\approx 800 \unit{MHz}$~away). We assume a dc field inhomogeneity of $1\unit{V/m}$, and demand the maximum dc Stark shift $\delta \omega < 0.1 \Omega$ during all microwave pulses (for these conditions, the maximum practical dipole moments could be achieved with $n\approx 30$, but higher $n$ still improves sensitivity due to longer $\tau_{\rm eff}$).

In pulsed-refocusing experiments, these parameters result in $S_{Fmin}\approx 3.9 \times 10^{-12} \unit{(V/m)^2 s}$. For noise at $1 \unit {MHz}$, this gives the figure of merit $\omega S_F(\omega)=4.2 \times 10^{-5}\unit{(V/m)^2}$, which is typically seen about $200 \unit{\mu m}$ away from the surface of non-cryogenic ion traps~\cite{Haffner_NJP.13.013032,Wineland_PRL.109.103001}.

To measure noise in this frequency range with spin-locking, the Rabi frequency will be lower (matching the noise frequency of interest), while the same requirement that dc Stark shifts must be small applies. Thus, $S_{Fmin}\approx 1.2 \times 10^{-8}-1.2 \times 10^{-10}\unit{(V/m)^2s}$ for the parameters listed above. Therefore, pulsed decoupling sequences, when practical, offer greater sensitivity when the atomic polarizability is limited by inhomogeneous dc electric fields. 

The dc field homogeneity requirement relaxes to $\Delta \omega \approx \Omega$ in a spin-lock sequence if the resonant $\pi/2$ pulses used to generate rotations from the $z-$axis to the equator are replaced by adiabatic sweeps of the microwave frequency. The Bloch vector would adiabatically follow the rotation axis in this case, minimizing the negative effects of Stark shift induced detuning. This would improve $S_{Fmin}\approx 1.2 \times 10^{-10}-1.2 \times 10^{-12}\unit{(V/m)^2s}$ when $\Omega \approx 2 \pi \times 1-10 \unit{MHz}$ for the parameters listed above. 

\subsection{Estimate of field noise near the surface}
\label{se:estimate}
Motivated by the approximate $1/f$ frequency scaling observed near microfabricated ion trap electrodes, we wish to determine the magnitude of the field noise near the chip surface assuming a noise power spectrum of the form $S_F(\omega)=S_0/|\omega|^\kappa$.

Coherence decay times near the chip surface during the spin-echo sequence shown in Fig.~\ref{fg:big_plots}(h) are $\tau=5.3 \pm 2 \unit{\mu s}$. Using the results in the Appendix, the decay time is consistent with $S_0=180 \unit{(V/m)^2}$ for $\kappa=1$, $S_0=9.6 \unit{(V/m)^2 s^{-1/4}}$ for $\kappa=3/4$, and $S_0=0.25 \unit{(V/m)^2 s^{-1/2}}$ for $\kappa=1/2$.

Spectroscopic measurements of the transition frequency may also be used to constrain the magnitude and frequency scaling of the noise. Near the chip, we observe Stark shifts of $400 \pm 100 \unit{kHz}$, consistent with $\langle F^2\rangle\approx 1300 \unit{(V/m)^2}$. If $S_F(\omega) = S_0/\omega^\kappa$, $\langle F^2\rangle \leq 2\int_{\omega_{ir}}^{\omega_c}d\omega \: S_0/\omega^\kappa$. The measurement sequence results in $\omega_{c} \approx 2 \times 10^5 \unit{s^{-1}}$ and $\omega_{ir} \approx 0.02 s^{-1}$. This constrains $S_0 \leq 40 \unit{(V/m)^2}$ for $\kappa=1$,$S_0 \leq 8 \unit{(V/m)^2 s^{-1/4}}$ for $\kappa=3/4$, and  $S_0 \leq 0.7 \unit{(V/m)^2 s^{-1/2}}$ for $\kappa=1/2$.

Thus we see that the spin-echo decay times and spectroscopic Stark shifts are inconsistent if we assume a model of $1/f^\kappa$ noise with $\kappa \geq 3/4$. The data are consistent with a noise frequency scaling of $1/f^{1/2}$, and with power spectral noise densities about six orders of magnitude larger than those observed near microfabricated ion traps at comparable distances \cite{Wineland_PRL.109.103001,Haffner_NJP.13.013032}. The observed noise is likely to have a technical origin, and measurements of spin-lock decay time as a function of Rabi frequency would serve to further constrain the frequency scaling of the noise as well as identifying any sharp features in the spectrum which may be associated with technical noise sources.


\section{Concluding remarks}

The focus of the present work has been the observation of coherent behavior of Rydberg atoms near the surface of an atom chip.  Spin-echo and spin-locking extend coherence times, and may be useful in conjunction with techniques for reducing Rydberg atom susceptibilty to low-frequency electric fields \cite{hyafil:2004,*mozley:2005,jones:2013}.  However, as discussed, coherent superpositions of Rydberg atoms may also be used to obtain estimates of  electric field spectral noise densities, particularly their frequency dependence, which is of relevance to anomalous ion-trap heating.  To use coherent superpositions of Rydberg atoms as field noise sensors, application of a dc field bias would improve noise measurement sensitivity and simplify the inference of spectral noise densities from spin-locking measurements.  However, as technical noise sources are reduced, and the sensitivity to noise improves, increasing attention will need to be paid to reducing the influence of the interactions between Rydberg atoms on these measurements.

\acknowledgments

We thank O. Cherry for atom chip fabrication and R. Mansour for use of the CIRFE facilities.  This work was supported by NSERC.

\appendix


\section{Quadratic coupling to fields with $1/f{^\kappa}$ spectral noise densities}
\subsection{Free evolution with quadratic coupling to noise}
In zero dc electric field, the leading order of the Stark shift is quadratic for low-angular momentum Rydberg states.  In general, a lack of first-order sensitivity to noise fields is desirable for maintaining coherence (consider, for example, the optimal working point of quantronium, where the biasing parameters are selected to eliminate first-order dependence~\cite{Vion_Sci.03052002}). 

When the leading order of the coupling of transition energy to a noise source $\lambda$ is quadratic, the decoherence rate due to noise depends on the statistics of $\lambda^2$, including not only the power spectral density $S_{\lambda^2}(\omega)$ but also higher orders in the statistics of $\lambda^2$~\cite{Makhlin_PRL.92.178301}. Taking into account the higher order statistics for non-Gaussian $\lambda^2$ (while assuming that $\lambda$ is Gaussian), the dephasing of Ramsey and Hahn spin echo experiments has been solved analytically in certain time regimes for pure $1/f$ noise, where $S_\lambda(\omega)=S_0/|\omega|$~\cite{Makhlin_PRL.92.178301,Ithier_PRB.72.134519}. It is useful to define 
\begin{equation}
\label{eq:Gamma_f}
\Gamma_f=\left(\frac{1}{2}\frac{\partial \omega_{if}^2}{\partial \lambda^2}\right)S_0.
\end{equation}
In the limit of $\Gamma_f t \ll 1$~\cite{Makhlin_PRL.92.178301,Ithier_PRB.72.134519}, 
\begin{equation}
\label{eq:short_times}
f_z(t) =\left[1 + \left(\frac{2}{\pi}\Gamma_f t \ln \frac{1}{\omega_{ir} t}\right)^2\right]^{-1/4},
\end{equation}
where $\omega_{ir}$ is a low-frequency (``infrared'') cut-off.
When $\Gamma_f t \gg 1$~\cite{Makhlin_PRL.92.178301,Ithier_PRB.72.134519},
\begin{equation}
\label{eq:long_times}
f_z(t) = e^{-\Gamma_f t /2}.
\end{equation}
At long times, the exponential time dependence and lack of dependence on $\omega_{ir}$ are surprising. Makhlin and Shnirman~\cite{Makhlin_PRL.92.178301} offer a qualitative explanation: the interaction between various low-frequency components due to the non-linear coupling of field noise to the phase effectively cuts off the $1/f$ behavior of the noise spectrum at frequencies below $\Gamma_f$. Below this frequency, the spectral density of the phase fluctuations is consistent with white noise rather than $1/f$ noise, which leads to exponential decay of the coherence at times $t\gg 1/\Gamma_f$~\cite{Makhlin_PRL.92.178301}.

\subsection{Monte Carlo results}

\label{se:1_over_f}
Analytical solutions for the decoherence $f_z(t)$ of Ramsey and spin-echo sequences may not be feasible for all couplings, time scales, and noise spectral density scalings of interest. In particular, analytical solutions in the case of quadratic coupling to noise with power spectral density of the form $1/f^\kappa$ only appear in the literature for $\kappa=1$ \cite{Makhlin_PRL.92.178301}.  However, Monte Carlo simulations averaging over many realizations of the noise provide another method of calculating $f_z(t)$ in cases of interest.  Here we present Monte Carlo coherence simulations over a range of $\kappa$, that are useful in Section \ref{se:estimate} of the main text.

For Ramsey or spin-echo sequences with quadratic coupling to noise, the coherence is~\cite{Ithier_PRB.72.134519}
\begin{equation}
f_z(t)=\left\langle \exp\left[ i \frac{1}{2}\frac{\partial \omega^2_{if}}{\partial \lambda^2}\int_0^t d\tau G(\tau)\lambda^2(\tau) \right] \right\rangle,
\end{equation}
with the gating function $G(\tau)$ alternating between $\pm 1$ at each refocusing $\pi$-pulse ($G(\tau)=1$ at all times for a Ramsey sequence).

We model random field noise as a sum of Fourier components with amplitudes determined by $S_\lambda(\omega)$ and random phases (we use $\lambda$ to represent the field noise). A time sequence of $2N$ discrete values of $\lambda(t)$ over a record length $T$ requires $N$ Fourier components. The frequencies of these components are evenly spaced and range from a lower cutoff $\omega_{ir}=2\pi/T$ to an upper cutoff $\omega_c=2\pi N/ T$. Summation of the Fourier components may be done efficiently with any fast Fourier transform algorithm. Averaging over $n$ realizations, the fractional uncertainty of $\langle \exp[i\delta \phi] \rangle$ is of the order $1/\sqrt{n}$.

In our simulations, we used $N=2^{22}\approx 4.2 \times 10^6$ Fourier components. We chose $\omega_{ir}=2 \pi \times 1 \unit{kHz}$, so that $\omega_c\approx 2\pi \times 4.2 \unit{GHz}$---decoherence times measured experimentally are in the $1-10 \unit{\mu s}$ range, so for time scales of interest we have $\omega_{ir} \ll 1/t \ll \omega_c$. We chose $\partial^2\omega_{if}/\partial \lambda^2=2 \pi \times 600 s^{-1}$ to match the magnitude of the polarizability difference between the $48s_{1/2}$ and $49s_{1/2}$ states (with the field magnitude in units of V/m). Modelled noise spectral densities were of the form $S_\lambda(\omega)=S_0/|\omega|^\kappa$, with convenient values of $S_0$ chosen at each $\kappa$.

\begin{figure}
\begin{center}
\includegraphics{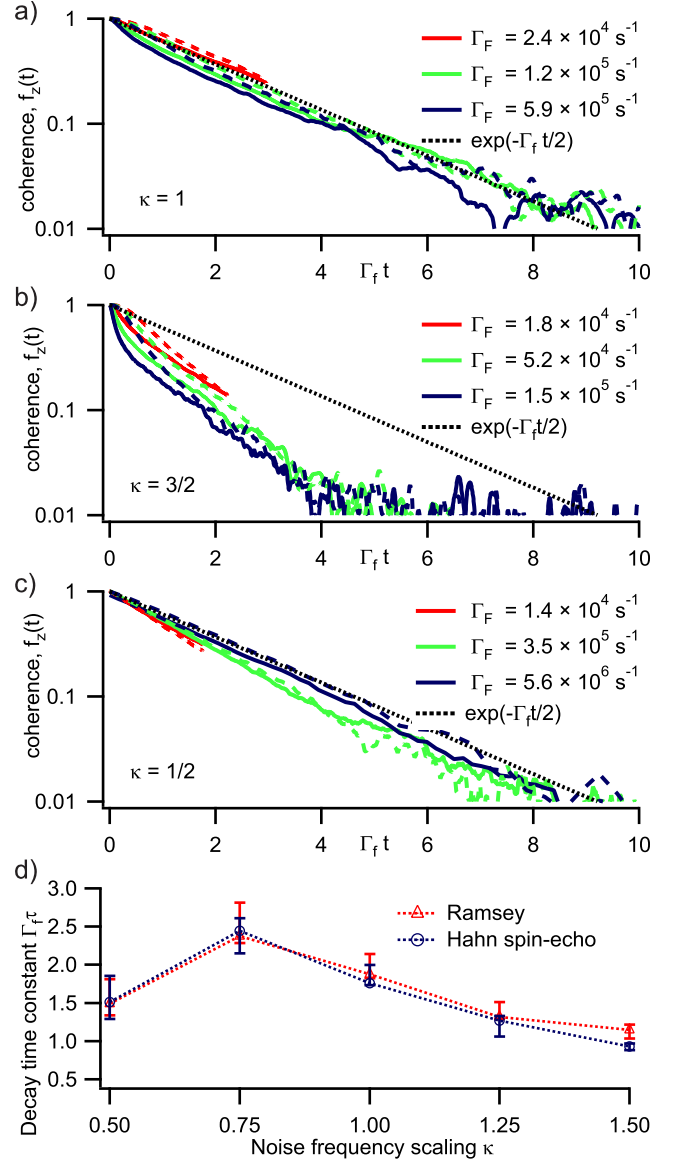}
\end{center}
\caption[Coherence decay with scaled time units in pure $1/f$ noise]{
(Color online)
\label{fg:hahn_ramsey}a)-c): Coherence decay for Ramsey sequences (solid lines) and Hahn spin-echo sequences (dashed lines) as a function of the scaled time $\Gamma_f t$. The power spectrum of the noise is of the form $S_\lambda(\omega)=S_0/\omega^\kappa$, with a)~$\kappa=1$, b)~$\kappa=3/2$, and c)~$\kappa=1/2$, with quadratic coupling to the energy level separation. Lower cutoff frequency $\omega_{ir}=2 \pi \times 1 \unit{kHz}$, fractional uncertainty of the calculated coherences is $1/\sqrt{n=10000} = 0.01$. d): time constant of the coherence decay, scaled to $\Gamma_f$, as a function of $\kappa$. The error bars show the variation of $\Gamma_f \tau$ at each $\kappa$ value as the noise amplitude $S_0$ is varied over about two orders of magnitude.}
\end{figure}

Results of the simulation for pure $1/f$ noise of the form $S_\lambda(\omega)=S_0/\omega$ are presented for various $S_0$ in Fig.~\ref{fg:hahn_ramsey}(a). It is convenient to plot time in units of the scaled time $\Gamma_f t$. At long times $t>1/\Gamma_f$, these simulations show identical exponential decay of coherence for Ramsey and Hahn spin-echo sequences, with $f_z(t)\propto e^{-\Gamma_f t/2}$, consistent with Eq.~\ref{eq:long_times}.
This behavior is independent of $\Gamma_f$, $\omega_{ir}$, and $\omega_c$ up to a prefactor (constant at long times) dictated by the decoherence at short times---this prefactor depends on the relative magnitudes of $\omega_{ir}$ and $\Gamma_f$. The $1/f$ noise falls off sufficiently fast with increasing frequency such that the results are independent of $\omega_c$ (provided that $t > 1/\omega_c$ for all times of interest). 

Analytical solutions for the decoherence during free evolution exist in the literature~\cite{Makhlin_PRL.92.178301,Ithier_PRB.72.134519}. However, reproducing the analytical result with the Monte Carlo simulations is useful for validating the model.  This is consistent with the picture of high-frequency noise (which cannot be effectively mitigated by refocusing) being the dominant source of decoherence in this regime, as expected from the exponential form of the decay at long times calculated analytically.

This exponential behavior at long times is not confined to pure $1/f$ noise. Simulations where the noise spectrum $S_\lambda(\omega)=S_0/\omega^\kappa$, with $1/2 \leq \kappa \leq 3/2$ show similar exponential time dependence of the coherence at sufficiently long times. This result suggests the possibility of a simple relationship between the noise amplitude and the decay rate of the coherence at long times. We extend the definition of $\Gamma_f$ beyond the $\kappa=1$ case:
\begin{equation}
\label{eq:new_Gamma_f}
\Gamma_f=\left(\frac{1}{2}\frac{\partial^2\omega_{if}}{\partial F^2}S_0 \right )^{1/\kappa},
\end{equation}
which is consistent with the previous definition of Eq.~\ref{eq:Gamma_f} when $\kappa=1$. 

For $\kappa=3/2$, coherence vs. $\Gamma_f t$ for various values of $\Gamma_f$ is plotted in Fig.~\ref{fg:hahn_ramsey}(b). The dynamic range in $S_0$ is similar to Fig.~\ref{fg:hahn_ramsey}(a); presumably the larger variation in the decoherence rate at short times is due to the larger amount of low-frequency noise present when $\kappa$ is larger.

The situation at long times is more complicated when $\kappa=1/2$, because the noise falls off more slowly with increasing frequency. Therefore, $f_z(t)$  depends on the relative magnitudes of $\omega_c$ and $\Gamma_f$. Coherence vs. $\Gamma_f t$ for various values of $\Gamma_f$ is plotted in Fig.~\ref{fg:hahn_ramsey}(c). The sensitivity to the exact nature of the high-frequency cutoff is fairly weak. At long times ($\Gamma_f t > 1$), the scaled time constant of exponential decay ranges from $\Gamma_f \tau = 1.29$ to $\Gamma_f \tau=1.85$ as $\Gamma_f$ is changed over two orders of magnitude while $\omega_c$ is kept constant. 

The scaled coherence decay time $\Gamma_f \tau$ at various values of $\kappa$ is plotted in Fig.~\ref{fg:hahn_ramsey}(d). The exponential behavior of the coherence decay for Ramsey and spin-echo sequences over a range of $\kappa$, with decay rates on the order $\Gamma_f$, suggests that the decoherence rate may be described by a relatively straightforward scaling law. In a way, this is an inconvenient result because the form of the coherence decay gives no information about the value of $\kappa$. However, if $\kappa$ can be determined through measurements with higher spectral selectivity (multi-pulse sequences, for example) then the decay rate at long times can easily be used to measure the overall noise amplitude (i.e.~$S_0$).

\end{document}